\documentclass[a4paper,fleqn,usenatbib,twocolumn]{aastex61}
\usepackage{newtxtext}
\usepackage{newtxmath}
\usepackage[T1]{fontenc}
\usepackage{ae,aecompl}
\usepackage{graphicx}
\usepackage{amsmath}
\usepackage{amssymb}
\usepackage{upgreek}

\begin{document}
\newcommand{\psr}{PSR\,J0250+5854}

\title{LOFAR discovery of a 23.5-second radio pulsar}

\author{C.\,M.\,Tan}
\affiliation{Jodrell Bank Centre for Astrophysics, School of Physics and Astronomy, University of Manchester, Manchester M13 9PL, UK}

\author{C.\,G.\,Bassa}
\affiliation{ASTRON, the Netherlands Institute for Radio Astronomy, Postbus 2, NL-7990 AA Dwingeloo, The Netherlands}

\author{S.\,Cooper}
\affiliation{Jodrell Bank Centre for Astrophysics, School of Physics and Astronomy, University of Manchester, Manchester M13 9PL, UK}
  
\author{T.\,J.\,Dijkema}
\affiliation{ASTRON, the Netherlands Institute for Radio Astronomy, Postbus 2, NL-7990 AA Dwingeloo, The Netherlands}

\author{P.\,Esposito}
\affiliation{Anton Pannekoek Institute for Astronomy, University of Amsterdam, Science Park 904, 1098 XH Amsterdam, The Netherlands}
\affiliation{INAF--Istituto di Astrofisica Spaziale e Fisica Cosmica - Milano, via E. Bassini 15, I-20133 Milano, Italy}

\author{J.\,W.\,T.\,Hessels}
\affiliation{ASTRON, the Netherlands Institute for Radio Astronomy, Postbus 2, NL-7990 AA Dwingeloo, The Netherlands}
\affiliation{Anton Pannekoek Institute for Astronomy, University of Amsterdam, Science Park 904, 1098 XH Amsterdam, The Netherlands}
  
\author{V.\,I.\,Kondratiev}
\affiliation{ASTRON, the Netherlands Institute for Radio Astronomy, Postbus 2, NL-7990 AA Dwingeloo, The Netherlands}
\affiliation{Astro Space Centre, Lebedev Physical Institute, Russian Academy of Sciences, Profsoyuznaya Str.\ 84/32, Moscow 117997, Russia}

\author{M.\,Kramer}
\affiliation{Max-Planck-Institut f\"ur Radioastronomie, Auf dem H\"ugel 69, D-53121 Bonn, Germany.}
\affiliation{Jodrell Bank Centre for Astrophysics, School of Physics and Astronomy, University of Manchester, Manchester M13 9PL, UK}

\author{D.\,Michilli}
\affiliation{Anton Pannekoek Institute for Astronomy, University of Amsterdam, Science Park 904, 1098 XH Amsterdam, The Netherlands}
\affiliation{ASTRON, the Netherlands Institute for Radio Astronomy, Postbus 2, NL-7990 AA Dwingeloo, The Netherlands}

\author{S.\,Sanidas}
\affiliation{Jodrell Bank Centre for Astrophysics, School of Physics and Astronomy, University of Manchester, Manchester M13 9PL, UK}

\author{T.\,W.\,Shimwell}
\affiliation{ASTRON, the Netherlands Institute for Radio Astronomy, Postbus 2, NL-7990 AA Dwingeloo, The Netherlands}

\author{B.\,W.\,Stappers}
\affiliation{Jodrell Bank Centre for Astrophysics, School of Physics and Astronomy, University of Manchester, Manchester M13 9PL, UK}

\author{J.\,van\,Leeuwen}
\affiliation{ASTRON, the Netherlands Institute for Radio Astronomy, Postbus 2, NL-7990 AA Dwingeloo, The Netherlands}
\affiliation{Anton Pannekoek Institute for Astronomy, University of Amsterdam, Science Park 904, 1098 XH Amsterdam, The Netherlands}

\author{I.\,Cognard}
\affiliation{Laboratoire de Physique et Chimie de l'Environnement et de l'Espace, Universit\'e d'Orl\'eans/CNRS, F-45071 Orl\'eans Cedex 02, France}
\affiliation{Station de Radioastronomie de Nan\c{c}ay, Observatoire de Paris, CNRS/INSU, F-18330 Nan\c{c}ay, France}

\author{J.-M.\,Grie{\ss}meier}
\affiliation{Laboratoire de Physique et Chimie de l'Environnement et de l'Espace, Universit\'e d'Orl\'eans/CNRS, F-45071 Orl\'eans Cedex 02, France}
\affiliation{Station de Radioastronomie de Nan\c{c}ay, Observatoire de Paris, CNRS/INSU, F-18330 Nan\c{c}ay, France}

\author{A.\,Karastergiou}
\affiliation{Oxford Astrophysics, Denys Wilkinson Building, Keble Road, Oxford OX1 3RH, UK}
\affiliation{Department of Physics and Astronomy, University of the Western Cape, Private Bag X17, Bellville 7535, South Africa}
\affiliation{Department of Physics and Electronics, Rhodes University, PO Box 94, Grahamstown 6140, South Africa}

\author{E.\,F.\,Keane}
\affiliation{SKA Organisation, Jodrell Bank Observatory, SK11 9DL, UK}

\author{C.\,Sobey}
\affiliation{International Centre for Radio Astronomy Research - Curtin University, GPO Box U1987, Perth, WA 6845, Australia}
\affiliation{CSIRO Astronomy and Space Science, PO Box 1130, Bentley WA 6102, Australia}

\author{P.\,Weltevrede}
\affiliation{Jodrell Bank Centre for Astrophysics, School of Physics and Astronomy, University of Manchester, Manchester M13 9PL, UK}

\correspondingauthor{C.\,M.\,Tan}
\email{chiamin.tan@postgrad.manchester.ac.uk}

\begin{abstract}
We present the discovery of {\psr}, a radio pulsar with a spin period of 23.5\,s. This is the slowest-spinning radio pulsar known. {\psr} was discovered by the LOFAR Tied-Array All-Sky Survey (LOTAAS), an all-Northern-sky survey for pulsars and fast transients at a central observing frequency of 135\,MHz. We subsequently detected pulsations from the pulsar in the interferometric images of the LOFAR Two-metre Sky Survey, allowing for sub-arcsecond localization. This, along with a pre-discovery detection 2 years prior, allowed us to measure the spin-period derivative to be $\dot{P}=2.7 \times 10^{-14}$\,s\,s$^{-1}$. The observed spin period derivative of {\psr} indicates a surface magnetic field strength, characteristic age and spin-down luminosity of $2.6 \times 10^{13}$G, $13.7$ Myr and $8.2 \times 10^{28}$ erg s$^{-1}$ respectively, for a dipolar magnetic field configuration. This also places the pulsar beyond the conventional pulsar death line, where radio emission is expected to cease. The spin period of {\psr} is similar to those of the high-energy-emitting magnetars and X-ray dim isolated neutron stars (XDINSs). However, the pulsar was not detected by the \emph{Swift}/XRT in the energy band of 0.3--10\,keV, placing a bolometric luminosity limit of $1.5 \times 10^{32}$\,erg\,s$^{-1}$ for an assumed $N_{\rm H}=1.35\times10^{21}$\,cm$^{-2}$ and a temperature of 85\,eV (typical of XDINSs). We discuss the implications of the discovery for models of the pulsar death line as well as the prospect of finding more similarly long-period pulsars, including the advantages provided by LOTAAS for this.
\end{abstract}

\keywords{pulsars: individual (\psr) -- stars: neutron -- radio continuum: general -- X-rays: individual ({\psr})}

\section{Introduction}

Rotation-powered pulsars are known to show a large range of spin periods, from the current fastest of 1.4\,ms~\citep{hrs+06} to 12.1\,s (Morello et al, in preparation). Most of these pulsars are detected through their radio pulsations, with a small fraction solely via high-energy emission~\cite[e.g.][]{sdz+10}. Many properties of pulsars, including their characteristic age and surface magnetic field strength, can be estimated through the measurement of their period ($P$) and period derivative ($\dot{P}$), assuming a dipole magnetic field. The period and its derivative are also used to define the location of the so-called ``death line'' on a $P$-$\dot{P}$ diagram, beyond which pulsars are no longer expected to emit coherent radio emission~\citep{rs75}. There are many different death line models~\cite[e.g.][see also references in Section~\ref{deathline}]{cr93,zhm00} with different dependences on the properties required for the generation of coherent radio emission. At their core is the prediction that we expect to observe very few, if any pulsars with spin periods greater than several seconds.

Searching for the slowest-spinning radio pulsars is thus motivated, in part, by constraining the emission mechanism. However, finding long-period pulsars ($>$5\,s) is a challenge in most pulsar surveys. Only five of the ten longest-period pulsars known have been found in periodicity searches, mainly due to their consistently larger aggregated flux over time~\citep[see v1.58 of the ATNF pulsar catalog\footnote{http://www.atnf.csiro.au/people/pulsar/psrcat};][]{mhth05}.~\citet{lbh+15} suggested that the lack of detections of long-period pulsars in the PALFA survey is primarily due to the presence of low-frequency `red' noise in the data. Furthermore,~\citet{hkr17} showed how the algorithms applied to remove this red noise also reduce the sensitivity towards long-period pulsars. Worse still, as most pulsar surveys have a relatively short dwell time of a few minutes per pointing, long-period pulsars could be missed because there are few or no pulses during the observation. Thus, many of the longest-period radio pulsars, with periods of several seconds, have been found through single-pulse searches instead and they are known as Rotating RAdio Transients~\citep[RRATs,][]{mll+06,km11}, but even they have a maximum period of 7.7\,s~\citep{kkl+11}.

Conversely, many of the long-period pulsars discovered are detected through emission in the high-energy regime. Magnetars are pulsars with spin periods in the range of 0.3$-$11.8\,s~\citep[with the possible exception of the 6.67 hours spin period 1E 161348$-$5055;][]{rbe+16} and surface magnetic field strength on the order of $10^{14}$\,G~\citep{kb17,eri18}. They are detected as bright, pulsed X-ray sources, with measured X-ray luminosities up to 10$^{36}$ erg s$^{-1}$. The observed X-ray luminosities from magnetars are generally larger than the expected rotational energy loss from the measured period derivatives, suggesting that the emission is ultimately powered by the decay of the strong magnetic field. Initially, the apparent lack of radio emission from magnetars was thought to be due to suppression of pair cascades required for coherent radio emission when the period derivative is larger than a critical value that depends on the period~\citep{bh98,bh01}. However, radio pulsations and magnetar-like properties have since been detected in 5 pulsars (XTE\,1810$-$197,~\citealt{crh+06}; 1E1547$-$5408,~\citealt{crhr07}; PSR\,J1622$-$4950,~\citealt{lbb+10}; PSR\,J1745$-$2900,~\citealt{sj13}; PSR\,J1119$-$6127,~\citealt{akts16,glk+16}).
 
Another class of high-energy-emitting, long-period pulsars are known as the X-ray Dim Isolated Neutron Stars~\citep[XDINSs,][]{hab07,tur09}. They are characterized by a soft, blackbody-like continuum X-ray emission, with temperatures ranging from 50--110\,eV, with no hard, non-thermal X-ray emission, and measured luminosities typically much lower than the magnetars. Only 7 XDINSs are currently known, with spin periods ranging from 3.4--11.3\,s.~\citet{vrp+13} computed evolutionary tracks showing that XDINSs could be old strongly-magnetized neutron stars and that some of them could actually descend from magnetars. Attempts to search for radio pulsations from XDINSs have so far resulted in non-detections~\citep{joh03,kvm+03,kml+09}.

Here we present {\psr}, a radio pulsar with a spin period of 23.5\,s, slower than any known radio pulsar, magnetar or XDINS. It was discovered using the LOw Frequency ARray~\citep[LOFAR;][]{hwg+13} as part of the LOFAR Tied-Array All-Sky Survey~\citep[LOTAAS\footnote{http://www.astron.nl/lotaas},][Sanidas et al., in preparation]{clh+14}.In Section~\ref{observation}, we describe the observation that led to the discovery of the pulsar, along with multi-wavelength follow-up observations. We describe the results in Section~\ref{result} and the implications of the discovery in Section~\ref{discussion}.

\section{Observations and analysis} \label{observation}
\subsection{LOFAR beamformed observations}
LOTAAS is an all-Northern-sky survey for pulsars and fast transients using LOFAR at a central observing frequency of 135\,MHz, with a bandwidth of 32\,MHz. The survey employs the 6 high-band antenna (HBA) stations of the inner core of LOFAR with a maximum baseline of 320\,m, known as the Superterp. Each 1-h pointing consists of three sub-array pointings (SAPs), centered at three nearby positions separated by 3.8 degrees. An incoherent beam is formed for each SAP with a full-width half-maximum (FWHM) of 5.5 degrees. At the center of each SAP, a hexagonal grid of 61 coherently added tied-array beams (TABs) is formed, each with a FWHM of 21$\arcmin$ at the central frequency and overlapping near the half-power point~\citep[see][for more details on LOFAR beamformed modes]{sha+11}. A further 12 TABs are placed within each SAP, either towards known sources outside the hexagonal grid, or at other predetermined positions. The Stokes I data are recorded with a sampling time of 492\,$\upmu$s and 2596 channels of 12.2\,kHz each.

The raw data are stored in the LOFAR Long Term Archive and then transferred to the Dutch National Supercomputer Cartesius where a \textsc{presto}-based~\citep{ran01,rem02} pulsar search pipeline is applied. Radio frequency interference (RFI) mitigation is performed on the data using~\textsc{rfifind} and then dedispersed using~\textsc{prepsubband} to dispersion measures ($\mathrm{DM}$s) from 0--550\,pc\,cm$^{-3}$, with step sizes of 0.01\,pc\,cm$^{-3}$ for $\mathrm{DM}$s below 40\,pc\,cm$^{-3}$, 0.05\,pc\,cm$^{-3}$ for $\mathrm{DM}$s between 40 and 130\,pc\,cm$^{-3}$ and 0.1\,pc\,cm$^{-3}$ otherwise. Each dedispersed time series is searched for pulsars with a Fast Fourier Transform (FFT)-based periodicity search with~\textsc{accelsearch}, but without searching for accelerated signals. A sifting algorithm that checks whether a candidate is detected in nearby $\mathrm{DM}$ values, as well as having a spectral significance higher than 5$\sigma$ is applied. The sifted candidates are then folded with~\textsc{prepfold}, yielding an average of 20,000 folded candidates per pointing. A machine learning classifier~\citep{tls+18} is then used to determine which of the folded candidates are more likely to be astrophysical in nature. A separate
machine learning classifier is also used to search the data for dispersed single pulses~\citep{mhl+18}.

\begin{figure*}
  \centering
  \includegraphics[width=0.8\linewidth]{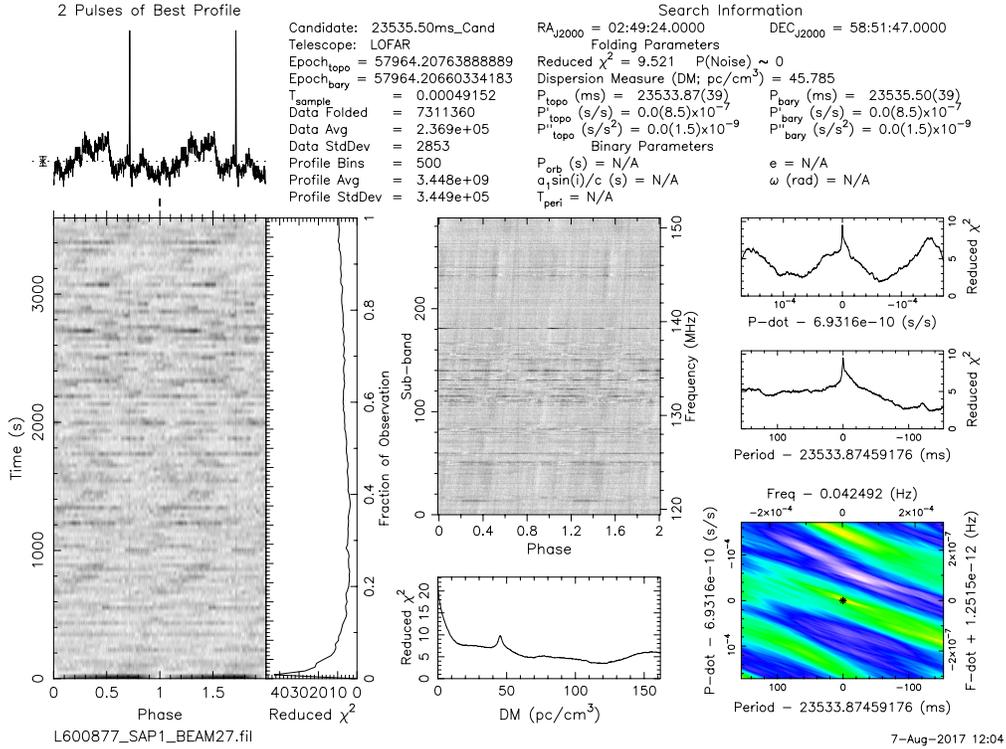}
  \caption{Discovery plot of {\psr}, folded at the fundamental period of 23.535 s, as inferred by the harmonically related candidates from the same observation. Note that, in the discovery observation, there was no candidate identified at the fundamental period itself.}
  \label{fig:discovery}
\end{figure*}

{\psr} was discovered in a LOTAAS observation obtained on 2017 July 30. For this observation the machine learning classifier reported several pulsar candidates at a $\mathrm{DM}$ near 45\,pc\,cm$^{-3}$ from a single TAB, with profile significance ranging from 8.4$\sigma$ down to 4.7$\sigma$. The significance of the weakest candidate is similar to some of the most significant candidates automatically classified as non-pulsar. Their spin periods were harmonically related, indicating a fundamental spin period of 23.535\,s. We verified that this was the true period by folding the data at this period, shown in Figure~\ref{fig:discovery}, as well as several harmonics. Single pulses from the pulsar are also detected in the TAB and these were used to confirm that this was the true period. We formed the time averaged profile of both the odd and even numbered single pulses using this period and the pulsar was seen in both. Folding at half the 23.535\,s period, however, resulted in the pulsed signal appearing in only the odd or even numbered average profile.

An initial follow-up observation was conducted using the discovery central frequency and bandwidth, but now using all 24 HBA stations of the LOFAR core, providing a longer maximum baseline of 2\,km, in order to localize the pulsar. It used a hexagonal grid of 127 TABs centered at the discovery position, each with a FWHM of $3\farcm8$. The pulsar was detected in two adjacent TABs resulting in a signal-to-noise ratio (S/N) weighted position of $\alpha_{\mathrm{J2000}}=02^{\mathrm{h}}50^{\mathrm{m}}11^{\mathrm{s}} \pm 4^{\mathrm{s}}$, $\delta_\mathrm{J2000}=+58\degr54\arcmin \pm 1\arcmin$. Another pair of follow-up observations centered on this new position, but using a separation between TABs of just $0\farcm5$ resulted in our best initial localization position of $\alpha_{\mathrm{J2000}}=02^{\mathrm{h}}50^{\mathrm{m}}17^{\mathrm{s}} \pm 1^{\mathrm{s}}$, $\delta_\mathrm{J2000}=+58\degr53\arcmin26\arcsec \pm 11\arcsec$. However, due to ionospheric beam jitter, these positions have an additional systematic uncertainty of up to $1\arcmin$.

Inspection of earlier LOTAAS pointings revealed pulsations from {\psr} in an observation obtained on 2015 August 5, with the TAB center located $8\arcmin$ from the position of the pulsar. The detection is much weaker than in the discovery observation, as this TAB is located further away from the SAP center compared to the discovery TAB from its SAP center, and as a result has a lower sensitivity. Hence, the pulsar was not detected by the search pipeline in this earlier observation. This fortuitous observation, taken almost exactly two years before the observation with which the pulsar was discovered, minimized any influence of position error on the period determination~\citep[see][]{lk05}. This, and the time span between detections, allowed us to initially estimate the spin-down rate of the pulsar to be $(3.5 \pm 1.4)\times10^{-14}$\,s s$^{-1}$.

Ten weekly timing observations of {\psr}, for an hour per epoch (corresponding to roughly 150 pulse periods) were obtained between 2017 September 9 and 2017 November 14 with the HBA antennas of the LOFAR core. Dual polarization, Nyquist sampled time-series in complex voltage (CV) mode were recorded for 400 sub-bands of 195.3\,kHz each, centered at 149\,MHz. The CV data were combined with the localization data obtained from the LOTAAS pointings and follow-up observations to obtain a timing solution for the pulsar.

A LOFAR Low Band Antenna (LBA) observation was made on 2017 September 23, using all 24 core LBA stations with 300 sub-bands of 195.3\,kHz each, centered at 62.4\,MHz. The data were recorded in CV mode. 

\subsection{LOFAR imaging}
\begin{figure}
  \includegraphics[width=\columnwidth]{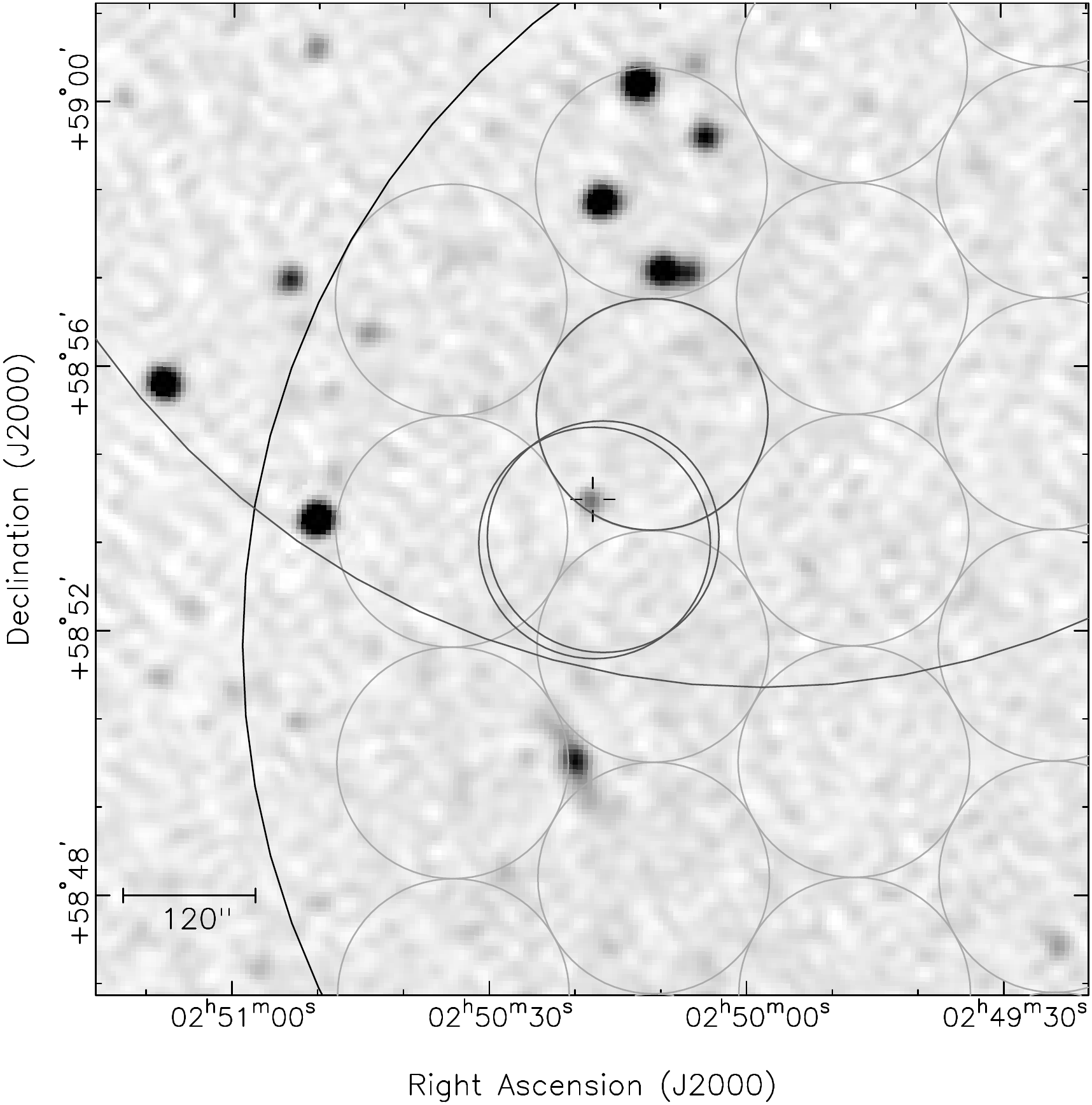}
  \caption{A $15\arcmin\times15\arcmin$ subsection of the LoTSS survey image over the 120 to 168\,MHz frequency range with $4\farcs5$ resolution. The position of \psr\ is indicated with tick marks ($10\arcsec$ in length). The large circles denote LOTAAS survey beams of $12\farcm3$ radius (at 119\,MHz) using the LOFAR Superterp stations, with the 2017 July 30 discovery observation shown in black, and the 2015 August 5 pre-discovery observation in dark gray. Beams from the LOTAAS confirmation observations, using the full LOFAR core, are shown with the light gray circles ($1\farcm75$ radius at 119\,MHz) and are laid out in the hexagonal pattern. \psr\ was detected in the dark gray circles, shown for three of these gridded confirmation observations.}
  \label{fig:lotss}
\end{figure}

The LOFAR Two-meter Sky Survey (LoTSS; \citealt{srb+17}) observed the region of \psr\ on 2016 August 31. The 8-h interferometric imaging observation was processed using a GRID implementation \citep{mod+17} of the standard direction-independent calibration pipeline\footnote{\url{https://github.com/lofar-astron/prefactor}} \citep{wwh+16,wwr+16}. The direction-dependent calibration and imaging was done using the LoTSS-DR1 processing pipeline (Shimwell et al., in prep.) which uses \textsc{KillMS} \citep{tass14,st15} to calculate direction-dependent calibration solutions and \textsc{DDFacet} to apply these during the imaging (\citealt{thm+17}). The LoTSS image (Figure~\ref{fig:lotss}) shows a single point-like radio source coincident with the best gridded position of {\psr}.

The LoTSS visibilities have sufficient time (1\,s sub-integrations) and frequency resolution (12.2\,kHz channels) to allow gated imaging on the \psr\ ephemeris. The ephemeris connecting the discovery, pre-discovery and confirmation observations was used to predict the pulse phase and spin period of \psr\ at the time of the LoTSS observation. Correcting for the 10\,s dispersion delay across the 120 to
168\,MHz LoTSS band, an \textit{on} image was made by selecting three 1\,s sub-integrations around the predicted arrival time of each pulse in the 8-h integration. The \textit{off} image used the remaining sub-integrations.

\begin{figure}
  \includegraphics[width=0.48\columnwidth]{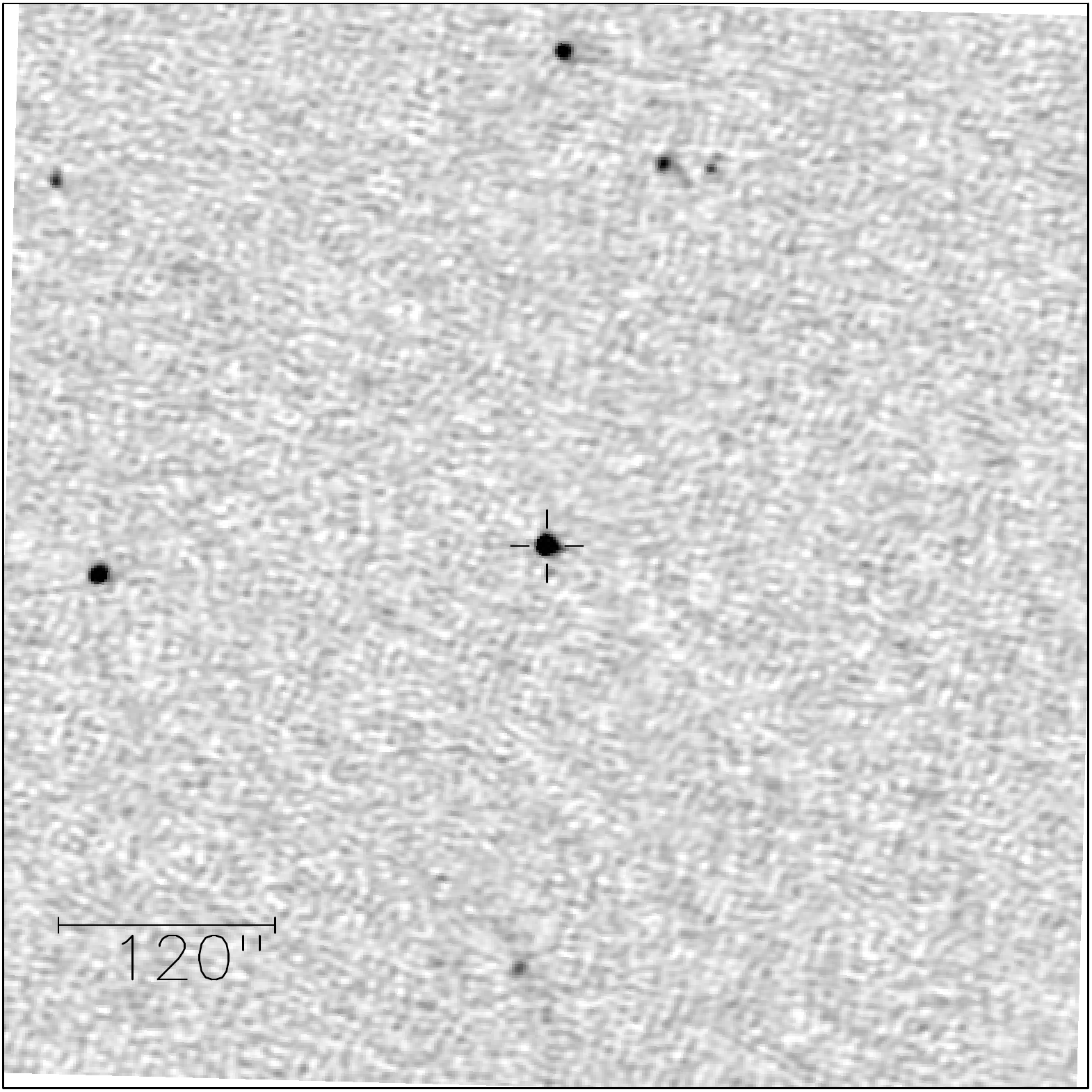}
  \includegraphics[width=0.48\columnwidth]{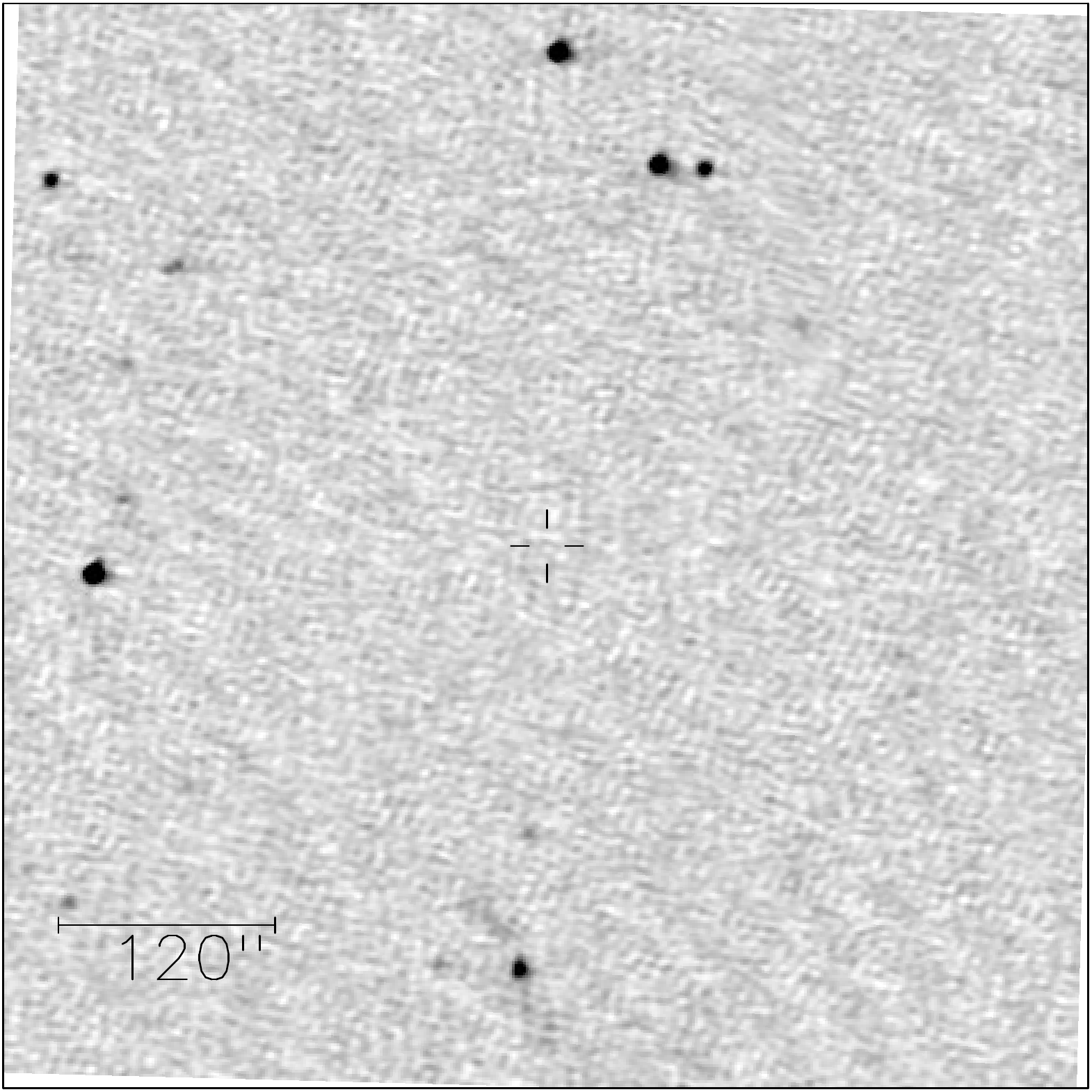}
  \caption{$10\arcmin\times10\arcmin$ subsections of the \textit{on} (left) and \textit{off} (right) LoTSS images, sampled at $1\farcs5$ spatial resolution.}
  \label{fig:onoff}
\end{figure}

The \textit{on}-\textit{off} images (Figure\,\ref{fig:onoff}) confirm that the point-like radio source is \psr, as the source is absent in the \textit{off}-image with a $5\sigma$ flux upper limit of 0.75\,mJy\,beam$^{-1}$. In the \textit{on}-image, \psr\ has a total gated flux density of 36.4\,mJy. As the \textit{on}-image uses 12.7\% of the 1-s sub-integrations of the 8-h integration, the pulse averaged flux density is 4.6\,mJy. The position of \psr\ in the \textit{on}-image is $\alpha_\mathrm{J2000}=02^\mathrm{h}50^\mathrm{m}17\fs781$, $\delta_\mathrm{J2000}=+58\degr54\arcmin01\farcs34$, which is consistent with the gridded TAB position described earlier.  In agreement with other fields processed using the LoTSS-DR1 pipeline (Shimwell et al., in prep.), the astrometric error is $0\farcs2$ and the flux scale is accurate to approximately 20\%.

\subsection{LOFAR Timing Analysis}
The weekly CV data from LOFAR HBA were coherently dedispersed and folded with \textsc{dspsr}~\citep{sb10}, while the LOTAAS data were incoherently dedispersed and folded. All the folded data were analyzed with~\textsc{psrchive}~\citep{hsm04}. The integrated pulse profiles were referenced against an analytic, noise-free template, obtained by fitting a single von Mises component to the profile with~\textsc{paas}, to measure pulse times-of-arrival (TOAs). A phase-connected timing solution was determined by modeling the TOAs with \textsc{tempo2}~\citep{ehm06,hem06}. An offset between the TOAs from the LOTAAS and CV data of 623$\pm$5\,${\upmu}$s is found and corrected by fitting a jump between the two sets of data\footnote{The offset comes from the fact that the LOTAAS data used a 2nd polyphase filter while the CV data do not.}.

\subsection{Green Bank Telescope Observation}
We observed {\psr} with the Green Bank Telescope (GBT) on 2017 October 25. The observation was 94 minutes long with 100\,MHz of bandwidth at a central observing frequency of 350\,MHz. The Stokes I data were recorded with the GUPPI backend~\citep{drd+08}. 

\subsection{Lovell Telescope Observations}
We observed {\psr} with the Lovell Telescope at Jodrell Bank, UK on 2 separate days; 2017 August 19 and 2017 November 30. The 1-hour long observations were recorded using the ROACH backend~\citep{kar11,bjk+16} with 384\,MHz of bandwidth at a central observing frequency of 1532\,MHz.

\subsection{Nan\c{c}ay Radio Telescope Observations}
We observed {\psr} with the Nan\c{c}ay Telescope on 2 separate days; 2018 July 7 and 2018 July 14 for 43 and 47 minutes, respectively. The observations were recorded using the NUPPI backend~\citep{ctg+13} with 512\,MHz of bandwidth at a central observing frequency of 1484\,MHz.

\subsection{X-ray Observations}

The field of {\psr} was observed for $\sim$600\,s with the Position-Sensitive Proportional Counter (PSPC) aboard \emph{ROSAT} during the All-Sky Survey (sequence rs930604n00, data collected between 1990 July and 1991 August; \citealt{bft+16}). No X-ray source was detected at the position of the pulsar. Following the prescriptions of \citet{bhi94} and using the X-ray image analysis package \textsc{ximage}\footnote{We used the command \textsc{uplimit}, adopting the Bayesian approach and the prior function described in \citealt{kbn91}, see the \textsc{ximage} user manual at \mbox{https://heasarc.gsfc.nasa.gov/xanadu/ximage/manual/ximage.html}.}, we set a 3$\sigma$ upper limit on the count rate in the 0.1--2.4\,keV band of $4.55\times10^{-2}$ counts\,s$^{-1}$.

Further observations of {\psr} were carried out with the \emph{Neil Gehrels Swift Observatory}'s X-Ray Telescope (XRT; \citealt{bhn+05}) in Photon Counting mode on 2018 March 23--24 (obs.ID 00010629001, net exposure: 5.8\,ks) and 2018 March 28 (obs.ID 00010629002, 4.0\,ks). We derived a 3$\sigma$ upper limit of $2.34\times10^{-3}$ counts\,s$^{-1}$ in the 0.3--10\,keV energy range from the non-detection of the combined data set.

We also inspected the stacked near-uv images of {\psr} obtained with the \emph{Swift} Ultraviolet/Optical Telescope (UVOT). We derived upper limits of 20.25 mag in the U filter (2.2\,ks exposure) and 21.21 mag in the UVW2 filter (7.4\,ks). As the UVOT data are less constraining than the XRT data, there is no further discussions of these observations below.

\section{Results} \label{result}
\subsection{Radio Observations}
\begin{figure}
  \includegraphics[width=\columnwidth]{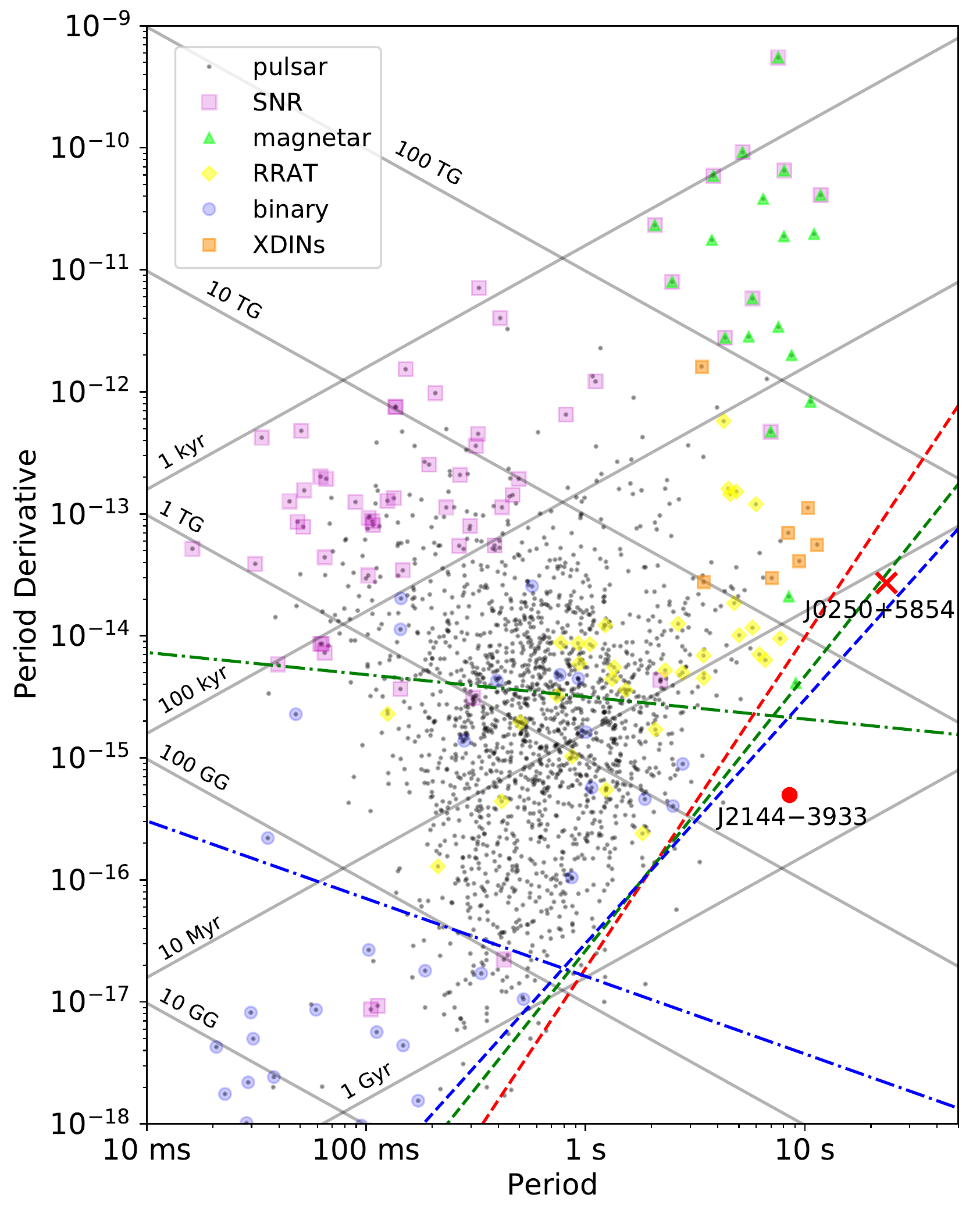}
  \caption{The $P$-$\dot{P}$-diagram of pulsars, characterizing them based on the measured spin period and spin period derivative. The plot is overlaid with lines indicating the characteristic age (1\,kyr, 100\,kyr, 10\,Myr, 1\,Gyr) and inferred surface magnetic field strength (10\,GG, 100\,GG, 1\,TG, 10\,TG, 100\,TG) of pulsars. Magnetars (green), XDINSs (orange), RRATs (yellow) and the 8.5\,s radio pulsar PSR\,J2144$-$3933 are indicated on the plot. {\psr} is located in a relatively empty part of the diagram. Several colored lines are plotted showing the various death line models based on pair productions, where pulsars below these lines are not expected to emit in radio. In red is the death line modeled by Equation 9 of~\cite{cr93}. The green and blue dashed lines are the death lines based on curvature radiation from vacuum gap and space-charged-limited flow (SCLF) models respectively, as proposed by~\cite{zhm00}. The green and blue dashed-dot lines are the death lines based on inverse Compton scattering from vacuum gap and SCLF models, also proposed by~\cite{zhm00}.}
  \label{ppdot}
\end{figure}

Using the LoTSS imaging position, we derived phase-connected timing solutions for {\psr} both with and without the inclusion of the 2015 pre-discovery data from LOTAAS. The solutions agreed with each other, indicating that the solution containing the 2015 point accounted for the correct integer number of pulsar rotations since that time. This solution is presented in Table~\ref{timingsolution}, with the position of the pulsar in the $P$-$\dot{P}$-diagram shown in Figure~\ref{ppdot}.

\begin{deluxetable}{lcc}
\tablehead{\colhead{Timing Parameters} & \colhead{Values}}
\tablecaption{The timing parameters of {\psr} obtained from the timing solution including the pre-discovery TOA from 2015 August 5. The parentheses indicate the 1-$\sigma$ uncertainty in the values. The position of the pulsar is fixed to the position obtained from LoTSS. The large reduced $\chi^2$ value obtained is likely due to each TOA being formed from a limited number of pulses, that could certainly add some jitter compared to the formal uncertainty.\label{timingsolution}}
\startdata
Right Ascension, $\alpha_\mathrm{J2000}$ & $02^{\mathrm{h}}50^{\mathrm{m}}17\fs78(3)$\\
Declination, $\delta_\mathrm{J2000}$ & $58\degr54\arcmin01\farcs3(2)$\\
Spin period (s) & 23.535378476(1)\\
Spin period derivative (s s$^{-1}$) & $2.716(7) \times 10^{-14}$\\
Dispersion Measure, $\mathrm{DM}$ (pc cm$^{-3}$) & 45.281(3)\\
Epoch of timing solution (MJD) & 57973\\
Solar system ephemeris model & DE405\\
Clock correction procedure & TT(TAI)\\
Time units & TCB\\
Timing span (MJD) & 57238.2-58071.9\\
Number of TOAs & 16\\
Weighted post-fit residual ($\upmu$s) & 493\\
Reduced $\chi^2$ value & 2.9\\
\hline
Derived Parameters &\\
\hline
Galactic longitude (deg) & 137.8\\
Galactic latitude (deg) & $-$0.5\\
$\mathrm{DM}$ distance (kpc) & 1.6\\
Characteristic age (Myr) & 13.7\\
Surface dipole magnetic field strength (G) & $2.6 \times 10^{13}$\\
Spin-down luminosity (erg s$^{-1}$) & $8.2 \times 10^{28}$\\
\enddata
\end{deluxetable}

We estimated the $\mathrm{DM}$ of {\psr} by measuring the TOA of pulses in five frequency sub-bands, with central frequencies of 117.7, 133.3, 148.9, 164.6 and 180.2\,MHz. The bandwidth of each sub-band is chosen to be 15.6\,MHz, in order to preserve sufficient S/N. To take into account possible profile evolution and scattering across the HBA band~\citep{hsh+12,bkk+16}, we first produced 5 different templates by fitting a single von Mises component to the profile of each sub-band of a single observation with \textsc{paas}. We then aligned the templates at the point corresponding to half of the peak height on the leading edge and used them to measure the TOA in each sub-band. The $\mathrm{DM}$ of the pulsar is then measured by minimizing the difference in these TOAs with \textsc{tempo2}. The measured $\mathrm{DM}$ of $45.281 \pm 0.003$\,pc\,cm$^{-3}$ gives a distance of approximately 1.6\,kpc, using Galactic electron density models from~\citet{cl02} and~\citet{ymw17}.

As there is some evidence for profile evolution we also attempted to refine the $\mathrm{DM}$ measurement by modeling the profiles, re-dedispersed at a $\mathrm{DM}$ of 45.281\,pc\,cm$^{-3}$, with two von Mises components. The resulting templates were aligned using two different methods, at the point corresponding to half of the peak height on the leading edge and at the peak of each template. The measured $\mathrm{DM}$ values were $45.262 \pm 0.003$ and $45.304 \pm 0.005$\,pc\,cm$^{-3}$ respectively. However, visual inspection of the pulse phase versus frequency plot for the lower DM shows some pulse broadening at the lowest frequencies, while at the higher value no significant difference was discernable. We therefore decided to use the data folded at $\mathrm{DM}$ of 45.281\,pc\,cm$^{-3}$ for further analysis.

\begin{figure}
  \includegraphics[width=\columnwidth]{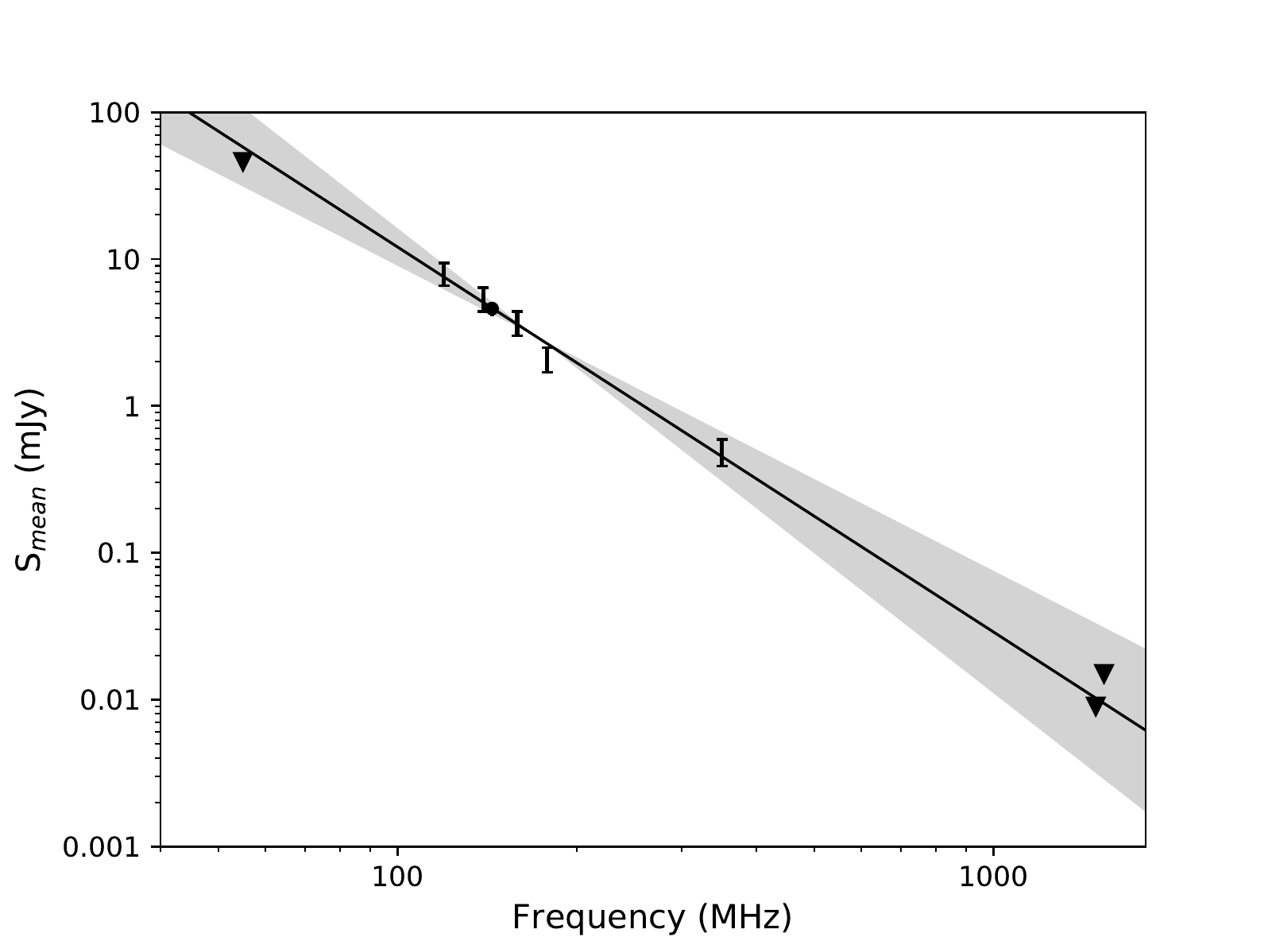}
  \caption{The fitted spectral index of {\psr} using the LOFAR timing and GBT observations. The shaded region is the 1-$\sigma$ uncertainty in the spectral index. The dot is the measured flux density by LoTSS. The triangles are the upper limit in flux densities obtained through the non-detections of LOFAR LBA, Nan\c{c}ay Radio Telescope and Lovell telescope observations respectively. The fitted spectral index is $-2.6 \pm 0.5$}
  \label{spectralindex}
\end{figure}

In order to study the spectrum of {\psr}, the LOFAR timing observations were flux-calibrated using the method detailed in~\cite{kvh+16}. Eight flux-calibrated observations were split into four frequency sub-bands with central frequencies of approximately 119.6, 139.1, 158.7 and 178.2\,MHz. The uncertainty in the average flux density of a single frequency sub-band is conservatively estimated to be 50\%~\citep{bkk+16}. The pulsar is also detected by the GBT, allowing us to estimate the flux density at 350\,MHz using the radiometer equation~\citep{lk05}, with a receiver temperature, T$_\mathrm{rec}$ of 23\,K, gain of 2\,K\,Jy$^{-1}$ and effective bandwidth of 60\,MHz due to band edge effects and the presence of RFI~\citep{slr+14}. The sky temperature T$_\mathrm{sky}$ in the direction of {\psr} is estimated to be 89\,K by extrapolating the value of T$_\mathrm{sky}$ at 408\,MHz~\citep{hssw82} with a spectral index of $-2.55$~\citep{lmop87,rr88}. The uncertainty of this flux density measurement is estimated at 20\%. The spectral index of the pulsar is modeled with a power law where $S_{\nu} \propto \nu^{\alpha}$. The result is shown in Table~\ref{tab:flux} and Figure~\ref{spectralindex}, with a fitted spectral index of $-2.6 \pm 0.5$ and $\chi^{2}_\mathrm{red}$ of 0.76.

\begin{deluxetable}{ccc}
\tablehead{\colhead{Frequency (MHz)} & \colhead{Flux Density (mJy)}}
\tablecaption{The flux densities of {\psr} at various observing frequencies. The parentheses indicate the 1-$\sigma$ uncertainty in the values. The flux density at 144\,MHz was obtained from the LoTSS image. An upper limit is placed for non-detections.\label{tab:flux}}
\startdata
55 & $<$46\\
119.6 & 8.0(14)\\
139.1 & 5.4(10)\\
144 & 4.6(9)\\
158.7 & 3.7(7)\\
178.2 & 2.1(4)\\
350 & 0.5(1)\\
1484 & $<$0.009\\
1532 & $<$0.015\\
\enddata
\end{deluxetable}

The fitted spectral index suggests that {\psr} has a steep spectrum compared to the average pulsar population~\citep[e.g.][]{blv13,bkk+16}. It also agrees with the flux density measurement from LoTSS. The pulsar is not detected with the LOFAR LBA nor the Lovell telescope and Nan\c{c}ay Radio Telescope. We estimated the upper limit on the flux densities using the radiometer equation and the derived system equivalent flux density of 27\,kJy in the most sensitive band of the LOFAR LBA between 50-60\,MHz~\citep{hwg+13} and assuming a detection threshold S/N of 10 and estimated duty cycle of 0.4\% based on the measured width of the profile at 129\,MHz. The upper limit on the flux density at 1532\,MHz from the Lovell telescope is also estimated using the radiometer equation with T$_\mathrm{rec}$ of 25\,K, T$_\mathrm{sky}$ of 5\,K, gain of 1\,K\,Jy$^{-1}$, bandwidth of 384\,MHz, estimated duty cycle of 0.3\% based on the measured width of the profile at 350\,MHz instead, as the pulse width is found to decrease at higher frequencies. and detection threshold S/N of 10. The upper limit on the flux density at 1484\,MHz from the Nan\c{c}ay Radio Telescope has T$_\mathrm{rec}$ of 35\,K, gain of 1.4\,K\,Jy$^{-1}$ and bandwidth of 512\,MHz, while the other values are the same as the 1532\,MHz limit. These limits are also shown in Table~\ref{tab:flux}. The upper limits are in slight tension with the spectrum inferred from detections between 100-400MHz. However, RFI strongly affects the LOFAR LBA data, and the high-frequency observations are possibly also affected by scintillation.

\begin{figure}
  \includegraphics[width=\columnwidth]{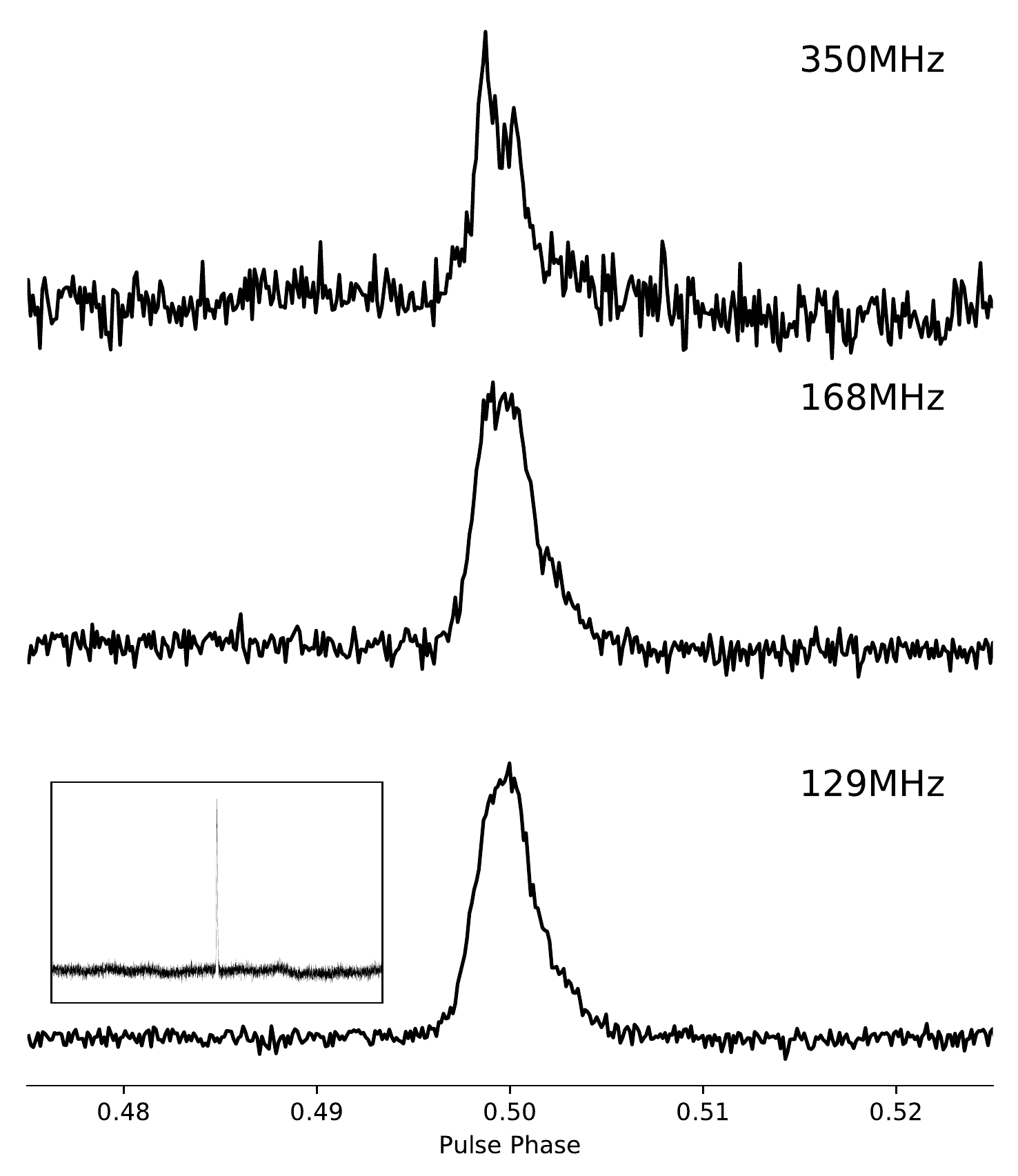}
  \caption{Integrated pulse profile of {\psr} at observing frequencies of 350 (GBT), 168 and 129\,MHz (LOFAR HBA), scaled to the same height for clearer illustration. The plot shows just 5\% of a full rotation.~\textit{Inset}: The pulse profile of {\psr} across the whole LOFAR HBA band seen over a full rotation.}
  \label{profevolution}
\end{figure}

Figure~\ref{profevolution} shows the integrated pulse profiles of {\psr} at central observing frequencies of 350, 168 and 129\,MHz respectively, after correcting for the dispersive delay. The pulsar exhibits significant profile evolution between 350\,MHz and the two LOFAR observing frequencies. The profile at 350\,MHz shows a double-peaked structure, with a stronger first peak. The pulsar shows similar pulse profiles in the two LOFAR observing bands: a single-peaked structure with a small trailing edge that resembles a scattering tail. However, the similarity between the pulse profiles suggests that this trailing edge is intrinsic to the pulsar instead of due to scattering, unless the frequency-dependent scattering time scale, $\tau$ of the pulsar has a power-law index much smaller than the theoretical predictions of $-$4 or $-$4.4~\citep[this has been seen in several pulsars with LOFAR, see][]{gkk+17}. We measured the profile widths at 129 and 168\,MHz by fitting templates of two von Mises functions to account for the trailing edge. The estimated pulse width at half-maximum, w$_{50}$ of the profiles at 129 and 168\,MHz, expressed in fraction of a full rotation, is $0.0035 \pm 0.0002$ and $0.0034 \pm 0.0002$ respectively, while the estimated pulse width at 10\% of maximum, w$_{10}$ of both profiles is $0.0075 \pm 0.0004$. The w$_{50}$ and w$_{10}$ of the profile at 350\,MHz is estimated to be $0.0026 \pm 0.0005$ and $0.008 \pm 0.001$ by fitting the profile with five von Mises functions, with a larger error due to uncertainty in the baseline of the profile.

\begin{figure}
  \includegraphics[width=\columnwidth]{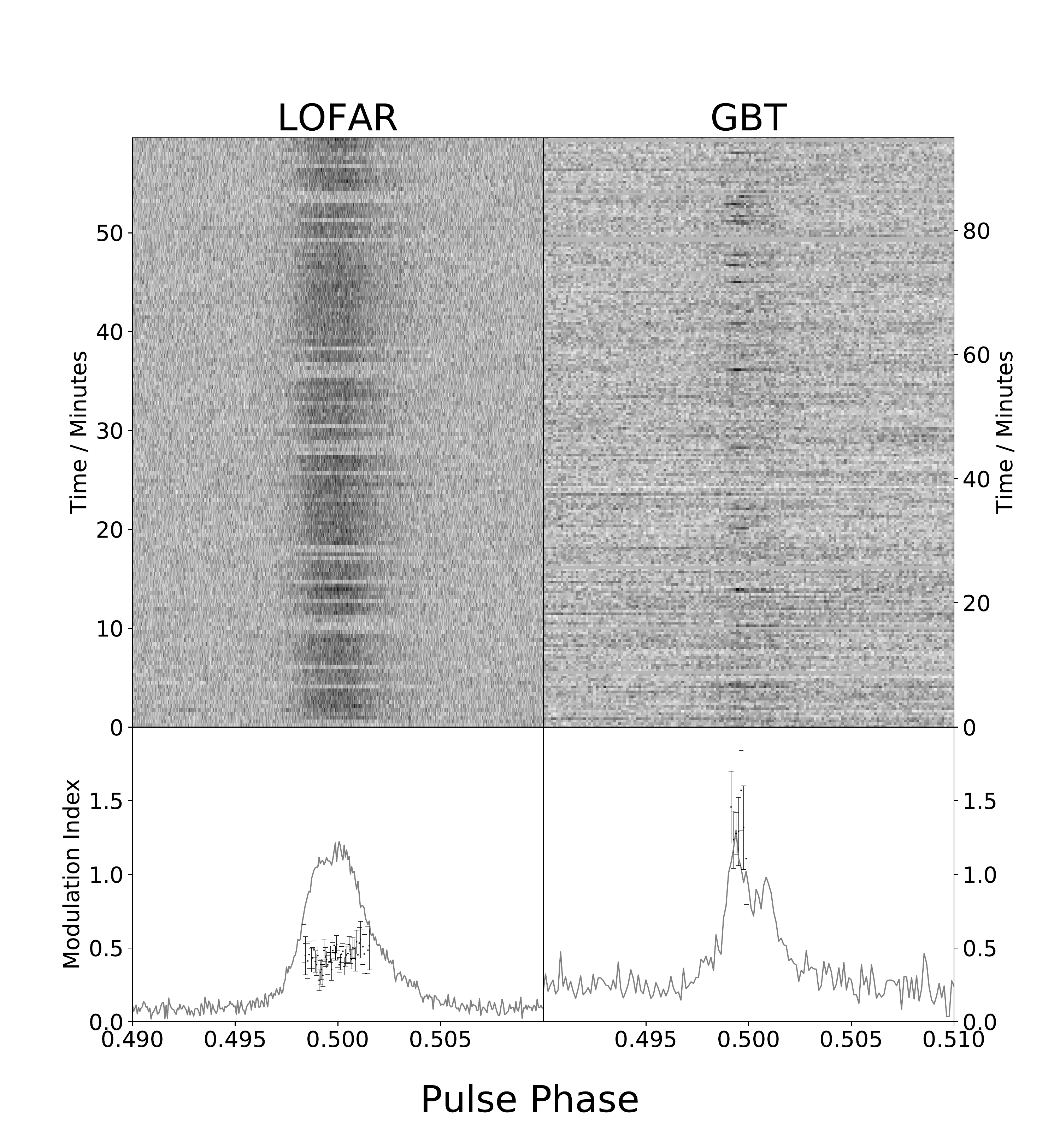}
  \caption{\textit{Top panels:} A side-by-side comparison of the single pulses of {\psr} from a 1-hour LOFAR observation and the GBT observation.~\textit{Bottom panels:} The modulation index of each phase bin of the on-pulse region on both observations, with the integrated pulse profiles
overlaid. Only phase bins with significant modulation are plotted.}
  \label{singlepulses}
\end{figure}

The single-pulse properties of {\psr} at LOFAR and GBT observing frequencies show significant differences (see Figure~\ref{singlepulses}). Over eight 1-h LOFAR observations, the pulsar is shown to switch off on timescales from one rotation up to about 5 minutes (12 rotations). We estimated the nulling fraction of each observation by measuring the pulse energy distribution using~\textsc{penergy} and~\textsc{pdist} of the~\textsc{psrsalsa} suite~\citep{wel16}. They show two clearly separable Gaussian distributions, corresponding to the pulsar being on and off. We estimated the nulling fraction to range from 9\% up to 42\% over the separate observations, with an average of 27\%. We calculated the modulation index~\citep{wel06} of each phase bin of the on-pulse region of a single LOFAR observation, to look for any variation in the single pulses. The modulation indices are found to be consistently around 0.4 across the on-pulse region, which could be due to the nulling. To confirm this, we removed the nulls from the pulse stack and recalculated the modulation indices. We found that only a few phase bins across the profile showed significant modulation, suggesting that there is no other variation in the single pulses apart from the nulling.

\begin{figure}
  \includegraphics[width=\columnwidth]{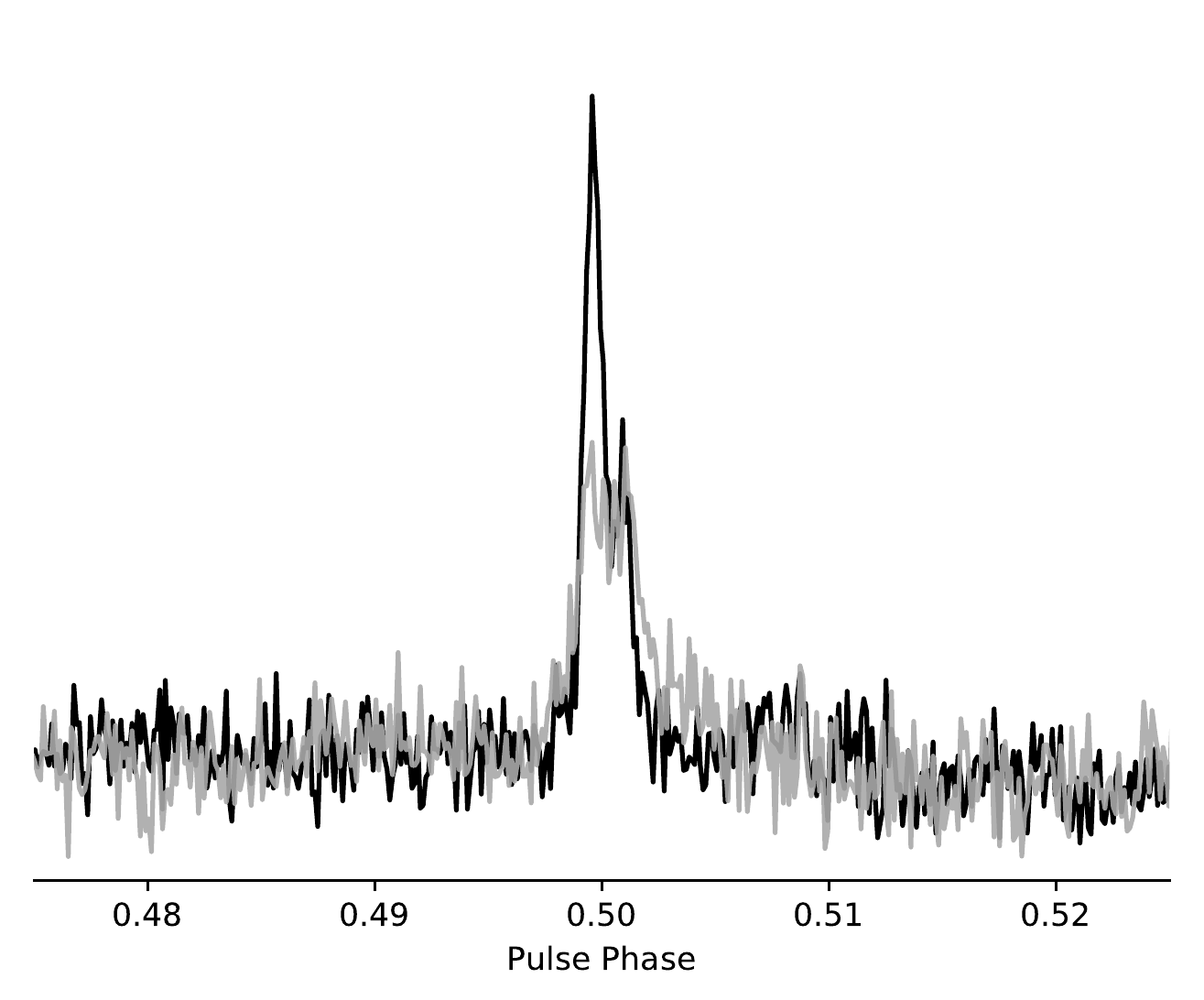}
  \caption{The integrated pulse profiles of the GBT observations consisting only of strong pulses (black) and of weak pulses (gray), overlaid on each other. The pulses were chosen through measurement of the pulse energy of the first component of the pulses. The profiles are normalized based on the off-pulse noise.}
  \label{splitprofiles}
\end{figure}

At 350\,MHz there appears to be a steady weak component with occasional highly variable strong pulses seen only in the first component of the integrated pulse profile. To confirm this, we calculated the modulation indices and found that there is indeed strong modulation only in the first component. We then measured the pulse energy distribution of the first component of the single pulses and found a tail of high energy pulses. We separate 22 of these strong pulses out of the 242 pulses based on the reported energy and produced two integrated profiles, consisting of the strong pulses, and the remainder, dubbed the weak pulses, as shown in Figure~\ref{splitprofiles}. These profiles suggest that {\psr} has a regular emission where the two components of the profile have roughly equal intensity. However, when the strong pulses are present, the first component becomes more prominent. We are unable to identify any possible nulling at this frequency due to the weak pulses having relatively low S/N.

\subsection{X-ray Observations}
The similarity in rotational parameters between {\psr} and XDINSs prompted a search for a possible high-energy counterpart. As the inferred spin-down luminosity $\dot{E}$ is too low for non-thermal emissions, we only considered possible thermal emissions for the X-ray analysis. 

The upper limit on the count rate of {\psr} from the PSPC is converted to a flux limit by using WebPIMMS\footnote{See \mbox{https://heasarc.gsfc.nasa.gov/cgi-bin/Tools/w3pimms/w3pimms.pl}.}, assuming a blackbody spectrum with temperature $kT=85$\,eV (approximately the temperature of the blackbody spectrum of XDINS RX\,J0720.4--3125; e.g. \citealt{vrp+13}) and a $\mathrm{DM}$-derived absorbing column $N_{\rm H}=1.35\times10^{21}$\,cm$^{-2}$ by assuming a 10\% ionization fraction in the interstellar medium~\citep{hnk13}. In the 0.1--2.4\,keV band, we obtained absorbed and unabsorbed limiting fluxes of $7.8\times10^{-13}$ and $6.5\times10^{-12}$\,erg\,cm$^{-2}$\,s$^{-1}$, corresponding to an upper limit on the luminosity of $2\times10^{33}$\,erg\,s$^{-1}$ at 1.6\,kpc. This limit is about an order of magnitude higher than the luminosity of RX\,J0720.4--3125, the brightest XDINS, and therefore does not constitute a severe limit with regard to the hypothesis of an enhanced thermal emission in {\psr}. 

\begin{figure}
  \includegraphics[width=\columnwidth]{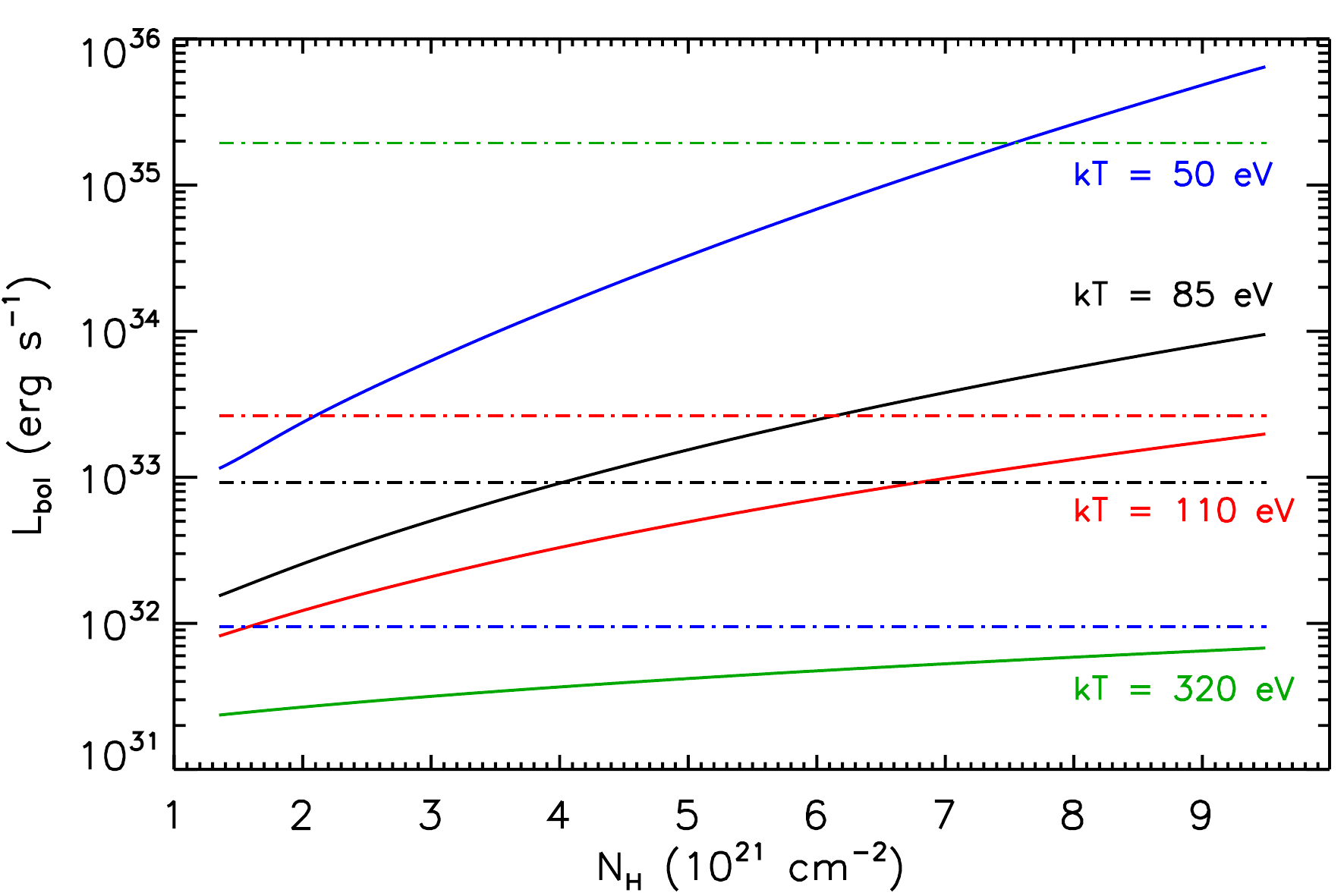}
  \caption{Upper limits on the bolometric luminosity of {\psr} derived from the \emph{Swift}/XRT data. The solid lines show the limits for different blackbody temperatures; the dotted-dashed lines indicate the bolometric luminosity corresponding to each temperature for an emitting region with radius of 12\,km.}
  \label{xuplims}
\end{figure}

For the non-detection of the dedicated \emph{Swift}/XRT observation, we explored a wider space of parameters. For the possible blackbody temperatures, we considered the values obtained by fitting the thermal components observed in RX\,J0720.4--3125 (85\,eV), as well as RX\,J2143.0+0654 (110\,eV) and RX\,J0420.0--5022 (50\,eV), the hottest and coolest XDINSs, respectively, and in the `low-magnetic-field' magnetar SGR\,0418+5729 (320\,eV; the values are from \citealt{vrp+13}). For the interstellar absorption, we take a range of values derived from the $\mathrm{DM}$, $N_{\rm H}=1.35\times10^{21}$\,cm$^{-2}$, to that predicted from the H\textsc{i} maps of \citet{dl90}, $\sim$$9\times10^{21}$\,cm$^{-2}$; other values derived from extinction and reddening~\citep{sfd98,sf11} are bounded by these limits. The different limits in the bolometric thermal luminosities, L$_\mathrm{bol}$ are shown in Figure\,\ref{xuplims}, together with the expected  L$_\mathrm{bol}$ from a radiating surface corresponding to a typical neutron star radius of 12\,km~\citep{lat17} at various temperatures.

\section{Discussion} \label{discussion}

The discovery of {\psr} has raised several questions regarding its unusually long spin period. In particular, we will discuss the likelihood of coherent radio emission from the pulsar based on current known models. We will also discuss the potential relation between {\psr} and the high-energy-emitting pulsars and the relationship between the spin period and the duty cycle of pulsars. Finally, we will discuss the prospects of detecting other pulsars of similarly long spin periods.

\subsection{The pulsar death line} \label{deathline}

The rotational parameters of {\psr} place it at the right end of the $P$-$\dot{P}$-diagram, shown in Figure~\ref{ppdot}, where few pulsars are known. Various death lines are presented in the diagram. As the exact mechanism of radio emission is still not well understood, these death lines are based on conditions in the pulsar magnetosphere required for pair cascade production, thought to be essential for the generation of radio emission.

One prevalent model for pair production is the vacuum gap model~\citep{rs75}, where there exists a vacuum gap on the polar cap of a neutron star in which the electric field and magnetic field are non-orthogonal. In order to sustain pair production, the potential difference across this gap must be sufficiently large. As a pulsar spin period increases, the thickness of the gap increases to maintain the potential difference. However, at large spin periods the increase in thickness is no longer enough to maintain the required potential difference. As a result, pair production and thus radio emission ceases. This period limit depends on the surface dipole magnetic field strength $B$, which can be inferred from the period derivative $\dot{P}$, and the curvature radius of the magnetic field lines, $r_{c}$. For a purely dipolar field, $r_{c} \sim 10^{8}$\,cm would require a potential difference that is too high to produce pairs even in most normal pulsars (see Equation 6 of~\citealt{cr93} for the death line of pulsars with pure dipole magnetic fields). Hence, strong non-dipolar fields are generally required near the surface.

\citet{cr93} defined a pulsar death valley bounded between the death line for a neutron star with a pure dipole magnetic field (their Equation 6), and the death line modeled by their Equation 9 (red dashed line in Figure~\ref{ppdot}) assuming $r_{c} \sim R = 10^{6}$\,cm, comparable to the radius $R$ of the neutron star, and a polar cap much smaller than possible with a pure dipole field. {\psr} is located beyond this death valley, suggesting an alternative model is needed to explain the emission.

\citet{cr93} also proposed a death line for an extreme case of twisted magnetic field lines at the polar cap (their Equation 10). {\psr} has not yet crossed this alternative death line. The presence of such a tangled field has been argued for in the bi-drifting behavior of PSR\,J0815+0939~\citep{sl17} by modeling the magnetosphere using a curvature radius $r_{c} \sim R = 10^{5}$\,cm.

\cite{zhm00} addressed the potential drop problem by taking into account relativistic frame-dragging effects~\citep{mt92,mh97} with a multi-pole magnetic field configuration. This places the death line (green dashed line, Figure~\ref{ppdot}) slightly to the left of the location of {\psr}.~\cite{gm01} argued that the death line for curvature radiation could shift further downwards by considering very curved magnetic field lines, with radii of curvature much smaller than the radius of a typical neutron star.~\citet{spi06} found a $(1+\mathrm{sin}^{2}~\alpha)^{-1/2}$ relationship between $B$ and the inclination angle $\alpha$ between the rotation and the magnetic axes of a pulsar, which reduces the maximum potential difference across the vacuum gap produced by pulsars of a given spin period by the same factor. This implies that emission from pulsars with large inclination angle would cease at smaller periods compared to those with small inclination angles.~\citet{zhm00} also considered the case where the radio emission originates from resonant inverse Compton scattering~\citep{zqlh97,hm98}, which places the pulsar death line (green dashed-dot line) below the location of {\psr}.

\cite{zhm00} also discussed the location of pulsar death lines within the framework of the space-charge-limited flow (SCLF) model~\citep{as79}. The death lines from curvature radiation and inverse Compton scattering (blue dashed and dash-dot lines, Figure~\ref{ppdot}) for this model, in the case of a multi-pole magnetic field, both lie below the location of~\psr.~\cite{hm11} suggest that the death line for curvature radiation in the SCLF model can be shifted downwards with a distorted magnetic field that produces an offset polar cap.~\cite{ztzw17} discussed the effects of different equations of state of a neutron star on the pulsar death line, in which heavier neutron stars could explain the presence of radio pulsars beyond the standard death line.

An alternative interpretation of the process underlying the cessation of pulsar emission is presented in~\cite{szmg14}. There, the radio emission, expressed as a fraction of the pulsar spin-down energy, is thought to have a maximum possible efficiency. A population synthesis model based on this assumption creates a ``death valley'' where in the published realization~\citep[Figure~5b in][]{szmg14}, long-period pulsars survive significantly past the traditional death lines. Using 1400\,MHz luminosities, a radio efficiency upper limit of 0.01 is derived: For {\psr}, we have determined a 10-$\sigma$ upper limit on the 1400\,MHz period-averaged flux density from the Lovell telescope of 15\,$\upmu$Jy. Using the 1.6\,kpc distance estimate from DM measurement, the radio luminosity as defined in Equation~2 of~\cite{szmg14} is $3.0 \times 10^{26}$\,erg\,s$^{-1}$. A comparison with the spin-down energy of $9 \times 10^{28}$\,erg\,s$^{-1}$ (Table~\ref{timingsolution}) produces a required efficiency for {\psr} of $<$0.004, below the 0.01 upper bound proposed by~\cite{szmg14} for successful generation of radio emission. This suggests that indeed, dim, long-period pulsars may continue to shine.

Nevertheless, the more recent death line models discussed above are able to explain the presence of {\psr}. While the pulsar is not as constraining as the 8.5\,s PSR J2144$-$3933 due to its large $\dot{P}$, it still rules out the conventional model described by Equation 9 of~\citet{cr93}. This also suggests that there could be a lot more pulsars in this region of the $P$-$\dot{P}$-diagram that are yet to be found.

\subsection{Relation to high-energy-emitting pulsars} \label{highenergy}

The similarity of the rotational parameters of {\psr} to the XDINSs and magnetars indicates a possible connection between them.~\citet{vrp+13} suggest that pulsars with rotational parameters similar to those of {\psr} could evolve from magnetars with an initial dipolar magnetic field strength on the order of 10$^{15}$G with a large decay in magnetic field. Using their Figure 10, we estimated the real age of {\psr} to be $\sim 3 \times 10^{5}$ yr, much smaller than the characteristic age, with an expected thermal luminosity of 10$^{34}$ erg s$^{-1}$ extrapolated from their Figure 11. However, the \emph{Swift}/XRT non-detection suggests that such an evolutionary track is unlikely, as it places the upper limit on $kT=85$\,eV for $N_{\rm H}=9\times10^{21}$\,cm$^{-2}$, which would require an emitting surface of radius 3 times larger than that of a canonical neutron star. PSR J0250+5854 does not show any evidence of a magnetar nature at the moment. Though transient, radio-loud magnetars have very different properties compared to PSR J0250+5854 (e.g. their spin-down luminosity is much larger and their characteristic age is much lower), it is still possible, however unlikely that it possesses a strong toroidal component and will manifest itself as a magnetar in the future.

The connection between {\psr} and XDINSs is still possible if XDINSs have a non-magnetar origin. The non-detection does not place a hard constraint as most of the known XDINSs have a lower L$_{\mathrm{bol}}$ than our most optimistic upper limit of L$_{\mathrm{bol}}=8 \times 10^{31}$ erg s$^{-1}$ for $kT=110$\,eV and $N_{\rm H}=1.35\times10^{21}$\,cm$^{-2}$. However, it would be difficult to detect soft X-ray emission (0.1-2.4\,keV) from {\psr} due to the large $N_{\rm H}$ along the line of sight to the pulsar, which absorbs a larger fraction of the soft X-ray flux.

Deep X-ray and optical observations of the 8.5\,s radio pulsar PSR\,J2144$-$3933 showed no evidence of enhanced thermal emission~\citep[L$_{\mathrm{bol}}$ > $\dot{E}$;][]{tmd+11}. It is possible that {\psr} also shows no such emission based on the non-detections.

\subsection{Spin period-duty cycle relationship of pulsars}
The small duty cycle of {\psr} is expected, given that the pulsar beam radius (for non-recycled pulsars), $\rho$, appears to scale as $\rho = K/\sqrt{P}$ (see e.g.~\citealt{lk05} and references therein). The factor $K$ depends weakly on frequency. At 400\,MHz, $K\sim 8$\,deg s$^{1/2}$ for a beam radius measured at 10\% of the peak intensity, $\rho_{10}$~\citep[e.g.][]{mr02}. For this pulsar, we expect $\rho_{10}\sim1.7$ deg. At a 10\% intensity level of the outer pulse edges, we measure a width of $w_{10} \sim 3$ deg. Therefore, within the uncertainties, $w_{10} \sim 2\rho_{10}$, and hence we can confirm the general expected scaling, which is based on the notion that with increasing light cylinder radius, the open field-line regions narrows, with a predominantly dipolar magnetic field structure as indicated by the $\sqrt{P}$ relationship. Note that the observed pulse width is determined at the height where radio emission leaves the magnetosphere, which is typically on the order of hundreds of kilometers~\citep{mr02,kg03}. This is far above the neutron star surface where pair production takes place. While the multipole components of the field are required to push the pulsar death line to longer periods (as argued in Section~\ref{deathline}), they are not necessarily significant anymore at the emission height given their rapid decay with distance. Hence at the emission height the magnetosphere is approximately dipolar.

The exact dependence between $w_{10}$ and $\rho_{10}$ depends on the geometrical configuration, i.e. the magnetic inclination angle $\alpha$ and the impact angle $\beta$~\citep[see e.g.][]{lk05}. Without polarization data measured over a sufficiently large duty cycle, it is difficult to determine these angles~\citep[e.g.~from the position angle swing in terms of a rotating vector model,][]{rc69}. Occasionally, it is possible to use statistical arguments to derive constraints~\citep[e.g.][]{kxl+98}, but we have failed in this particular case, as a wide range of $\alpha-\beta$ combinations are consistent with the observed width and inferred beam radius, given, in particular, the uncertainty in the $K$ factor when scaling to such large periods.

Nevertheless, it is clear that overall the beam radius is small. The corresponding beaming fraction, $f$, (i.e.~the fraction of celestial sphere covered by the sweeping beam) is also small, reducing the likelihood to detect such long-period pulsars for randomly oriented beam directions. Assuming a random distribution in $\alpha$ between 0 and 180 deg, one finds an expectation value for $f=(1-\cos\rho)+(\pi/2-\rho)\sin\rho$~\citep{ec89}, where $\rho$ is expressed in radians. Inserting our inferred $\rho_{10}$, this implies $f\sim 0.05$, so only 5\% of such long-period pulsars are observable, and the number of potentially existing pulsars is correspondingly larger. However, if $\alpha$ is small, for instance due to an alignment of the pulsar magnetic axis with age~\citep[e.g.][]{jk17}, then the beaming fraction is much smaller still, and even more such pulsars could exist. Other effects may also alter the effective beam radius, such as unfilled beams~\citep{lm88} or different emission heights as seen in different pulsars~\citep{lk05}. Overall, we find it remarkable that the observed pulse width essentially follows the extrapolation from smaller periods.

The period dependence of $f$ as shown by {\psr} is important in pulsar population synthesis for explaining why pulsars do not pile up near the death line~\citep{hbwv97,ls10,jk17}, without needing to invoke an alternative explanation of magnetic field decay in non-recycled, isolated pulsars~\citep[cf.][]{lv04,jk17}. Furthermore, the probable non-magnetar origin (as discussed in Section~\ref{highenergy}) and determination of a $3 \times 10^{13}$\,G field in a 23.5\,s pulsar (Table~\ref{timingsolution}) argues against magnetic field decay.

\subsection{Measuring the dispersion measure}
Our inability to obtain a definitive measurement of $\mathrm{DM}$ for {\psr} highlights the difficulty in disentangling the pulse shape evolution at the large fractional bandwidth and low frequencies of LOFAR~\citep[cf.][]{hsh+12}. The different templates produced for each of the sub-bands of the pulsar suggest that there is profile evolution across the bandwidth of the LOFAR observation. However, the evolution does not look like scattering as the width of the templates does not increase at lower frequencies and in fact we seem to lose a component seen at higher frequencies. The frequency evolution of the profiles also means that there is no obvious reference point in aligning the templates to measure the $\mathrm{DM}$. For example, when we align the templates at the peaks, the TOAs generated at 180.2\,MHz do not line up with others after fitting for $\mathrm{DM}$.

\subsection{Detectability of long-period pulsars}
We have shown that more pulsars with similarly long period to {\psr} could exist, and if they are detectable, will likely have small duty cycles as well. This would make FFT-based periodicity searches less sensitive to these sources. Hence, many of the longer period pulsars discovered recently by various surveys~\citep{kle+10,kkl+15,dsm+16} are first detected via single pulses as RRATs, with the underlying periodicity only detected with longer follow-up observations. However, {\psr} was discovered through an FFT-based periodicity search. This is primarily due to the relatively consistent flux from the pulsar apart from the nulling, combined with the 1-hour-long observations of LOTAAS giving us a large number of pulses ($\sim150$) for detection. For comparison, the two-minute dwell time of the GBNCC survey~\citep{slr+14} would only see 5-6 pulses. It should be noted that the pulsar was only detected at the fifth and higher harmonics in the discovery observation. While the fifth harmonic has a detection significance of 9$\sigma$ as reported by \textsc{accelsearch}, the lower harmonics are not detected down to significance of 2$\sigma$. The FFT-based periodicity search of LOTAAS is restricted to a minimum Fourier domain frequency of 1/16\,Hz and a maximum number of harmonics of 16. This meant that the fundamental frequency of the pulsar will not be detected by the search pipeline. The lower harmonics of the pulsar were not detected partly due to the presence of red noise in the low frequency regime. The small duty cycle of the pulsar also meant that the power of the pulsar in the Fourier domain is spread over a large number of harmonics, thus requiring the harmonic summing in the Fourier spectrum of a large number of harmonics for the pulsar to be detectable~\citep{mt77}. Fortunately, the large flux density of the pulsar at LOTAAS observing frequencies resulted in large power of the higher harmonics, allowing us to detect the pulsar through the summation of higher harmonics.

The discovery of {\psr} was also aided by the fact that the S/N shows a clear and narrow peak in $\mathrm{DM}$ space (the peak is only 4 pc cm$^{-1}$ wide). The pulsar is thus well distinguished from zero-$\mathrm{DM}$ RFI, which would be more difficult at higher observing frequencies. The discovery was also helped by the steep spectrum of {\psr}. If other long-period pulsars showed similar spectral indices then they would be easier to detect with LOFAR as well.

Our observation of {\psr} at a frequency of 350\,MHz showed somewhat different emission properties compared to the LOFAR observations. The pulsar showed sporadic strong single pulses at 350\,MHz. This suggests that the emission mechanism is strongly frequency dependent and perhaps results in less stable emission at higher frequencies. It is possible that there are other long-period pulsars that show variable single pulses at higher observing frequencies, which are more likely to be discovered through single pulse searches. On the other hand, they might exhibit more regular emission at lower frequencies that would require periodicity searches and a long dwell time that LOTAAS provided to be detected.

The discovery of {\psr} has significantly expanded the known range of rotation-powered pulsar periods. However, FFT-based periodicity searches will likely miss them if they have lower flux compared to {\psr}. The presence of red noise would also reduce the sensitivity towards long-period pulsars. Furthermore, if {\psr} had a larger duty cycle, the pulsar would not have been detectable, due to the smaller number of significant harmonics. Hence it is likely that we are missing out on more long-period pulsars due to these shortcomings of FFT-based periodicity searches. Recently, there has been renewed interest in using the Fast Folding Algorithm~\cite[FFA,][]{sta69} to search for pulsars. Various different implementations of FFA have been developed or are currently in preparation~\cite[][Morello et al. in preparation]{kml+09,cbc+17,pkr+18} with the expectation that FFA will be more sensitive towards long-period pulsars. In fact, in a preliminary analysis, the FFA implementation of Morello et al. (in preparation) detected {\psr} at the fundamental period with a high S/N of 40, much higher than the significance of the detection made by the FFT. The addition of FFA to the processing pipeline of LOTAAS is ongoing, and it will potentially discover new pulsars that will populate the area in the $P$-$\dot{P}$-diagram around {\psr}.

\acknowledgments
This paper is based (in part) on data obtained with the International LOFAR Telescope (ILT) under project codes LT5\_004 and DDT8\_004. LOFAR~\citep{hwg+13} is the Low Frequency Array designed and constructed by ASTRON. It has observing, data processing, and data storage facilities in several countries, that are owned by various parties (each with their own funding sources), and that are collectively operated by the ILT foundation under a joint scientific policy. The ILT resources have benefited from the following recent major funding sources: CNRS-INSU, Observatoire de Paris and Universit\'e d'Orl\'eans, France; BMBF, MIWF-NRW, MPG, Germany; Science Foundation Ireland (SFI), Department of Business, Enterprise and Innovation (DBEI), Ireland; NWO, The Netherlands; The Science and Technology Facilities Council, UK. The Nan\c{c}ay Radio Observatory is operated by the Paris Observatory, associated with the French Centre National de la Recherche Scientifique (CNRS) and Universit\'{e} d'Orl\'{e}ans. This work used data obtained with The Neil Gehrels Swift Observatory, which is a NASA mission with participation of the Italian Space Agency and the UK Space Agency. We thank the Swift PI Brad Cenko, the duty scientists and science planners for making the observations possible. The research leading to these results has received funding from the European Research Council under the European Union's Seventh Framework Programme (FP7/2007-2013) / ERC grant agreement nr. 337062 (DRAGNET; PI: Hessels). This work was carried out on the Dutch national e-infrastructure with the support of SURF Cooperative. Computing time was provided by NWO Physical Science. We thank Vincent Morello for testing the detectability of PSR J0250+5854 using his FFA implementation. JWTH acknowledges funding from an NWO Vidi fellowship. PE acknowledges funding in the framework of the NWO Vidi award A.2320.0076 and of the ASI--INAF contract `ULTraS' (N. 2017- 14-H.0).
\software{\textsc{presto} \citep{ran01,rem02}, \textsc{dspsr} \citep{sb10}, \textsc{psrchive} \citep{hsm04}, \textsc{tempo2} \citep{ehm06, hem06},  FTOOLS (v6.21; \citealt{bla95})}

\bibliographystyle{aasjournal.bst} \bibliography{references.bib}

\end{document}